\documentstyle[epsf,graphicx,rotating]{mn2e}
\input{psfig}

\newdimen\digitwidth
\setbox0=\hbox{\rm0}
\digitwidth=\wd0
\catcode `@=\active
\def@{\kern\digitwidth}

\begin{document}
\def\la{\mathrel{\hbox{\rlap{\hbox{\lower4pt\hbox{$\sim$}}}\hbox{$<$}}}}
\def\ga{\mathrel{\hbox{\rlap{\hbox{\lower4pt\hbox{$\sim$}}}\hbox{$>$}}}}
\def\proptosim{\mathrel{\hbox{\rlap{\hbox{\lower4pt\hbox{$\sim$}}}\hbox{$\propto$}}}}
\font\sevenrm=cmr7
\def\OII{[O~{\sevenrm II}]}

\title{The Disc-Jet Relation in Strong-lined Blazars}

\author[V. D'Elia, P. Padovani \& H. Landt]{Valerio D'Elia$^{1,2,3}$, Paolo Padovani$^{1,4}$,
Hermine Landt$^{1,5}$\\
$^1$ Space Telescope Science Institute, 3700 San Martin Drive, Baltimore, MD 21218, USA\\
$^2$ Dipartimento di Fisica, II Universit\`a di Roma ``Tor Vergata'', Via della Ricerca Scientifica 1, I-00133 Roma, Italy\\
$^3$ Osservatorio Astronomico di Roma, Via di Frascati 33, I-00040 
Monteporzio, Italy (current address)\\ 
$^4$ ESA Space Telescope Division\\
$^5$ Hamburger Sternwarte, Gojenbergsweg 112, D-21029 Hamburg, Germany}

\date{Accepted~~, Received~~}

\maketitle

\begin{abstract}
The relation between accretion disc (thermal emission) and jet (non-thermal
emission) in blazars is still a mystery as, typically, the beamed jet emission
swamps the disc even in the ultraviolet band where disc emission peaks. In
this paper we estimate the accretion disc component for 136 flat-spectrum
radio quasars selected from the Deep X-ray Radio Blazar Survey. We do this by
deriving the accretion disc spectrum from the mass and accretion rate onto the
central black hole for each object, estimated using the emission line widths
and the power emitted from the broad line region. We find that non-thermal
emission dominates the optical/UV band of our sources. The thermal component,
in fact, is, on average, $\sim 15$ per cent of the total and $\ga 90$ per cent
of the objects in the sample have a thermal component $< 0.5$ of the total
luminosity. We then estimate the integrated disc and kinetic jet powers 
and find
that, on average, the disc luminosity is $\sim 1$ to 20 times the jet power
(depending on the uncertainties in the estimation of the latter quantity). A
comparison with previous, independent results favours a scenario in which jet
and disk powers are of the same order of magnitude. Extraction of energy from
a rotating black hole via the ``Blandford-Znajek'' mechanism fails to explain
the estimated jet power in the majority of our sources. Finally, we find that
the typical masses for our sources are $\sim 5 \times 10^8 M_{\odot}$ and
that, contrary to previous claims, about one quarter of our radio quasars have
relatively small ($< 3 \times 10^8 M_{\odot}$) black hole mass.
\end{abstract}


\section{Introduction}
Blazars are probably the most extreme class of Active Galactic Nuclei (AGN)
due to their peculiar features: strong, high-frequency, flat radio emission
from compact cores, often expanding at superluminal speed (e.g., Jorstad et
al. 2001); irregular, rapid variability (e.g., Boettcher 1999); strong radio
and optical polarization (e.g., Yuan et al. 2001); and a broad continuum
extending to the GeV energies and beyond (e.g., Mukherjee 2001).

The defining features of the blazar class can be explained within the scenario
first proposed by Blandford \& Rees (1978). Namely, radiation from these
sources is likely produced in a relativistic jet of particles with bulk
Lorentz factor $\Gamma \sim 10 - 20$; when the jet is observed at small angle,
the so-called ``beaming'' effect is responsible for most of the above
features.  The spectral energy distribution (SED) of blazars commonly presents
a double peak structure (Fossati et al. 1998). The first peak (usually located
between the infrared and the UV band) is interpreted as synchrotron emission
from the relativistic electrons in the jet. The same electron population can
inverse Compton scatter the synchrotron produced photons (synchrotron self
Compton) and photons from the disc or the emission line regions (external
Compton) to produce high energy radiation peaking in the X-ray to $\gamma$-ray
region (Ghisellini 1999).

Among the blazar class, two subclasses can be identified (see Urry \& Padovani
1995). The first one is constituted by the Flat Spectrum Radio Quasars
(FSRQ). These are very powerful sources (exceeding in some cases $L \sim
10^{47}$ erg s$^{-1}$) which show strong and broad emission lines. To the
second one belong the BL Lacertae objects (BL Lacs), less luminous, and
lacking emission lines with equivalent width $ \ge 5$ \AA.

Due to the strong and relativistically amplified emission from the jet, the
thermal contribution from the accretion disc to the total luminosity cannot be
in general directly observed in blazars. For this reason the relation between
disc and jet is still uncertain for this class of sources (see Celotti,
Padovani \& Ghisellini 1997 and references therein).

In this paper we try to disentangle the blazar disc and jet components in the
optical/UV band. We do this by evaluating masses and accretion rates for a
sample of blazars, using these parameters to compute an accretion disc
emission model. To this order, information on the accretion disc continuum is
required, but that is generally swamped by the beamed jet emission. Therefore,
we use the Broad Line Region (BLR) luminosity and relate it to the thermal
emission. Thus, we are led to take in account in our analysis only FSRQ
sources, which show strong emission lines (although some BL Lacs show some 
emission lines, these sources are thought to have a very weak, if any,
accretion disc component; e.g., Cavaliere \& D'Elia 2002).

The paper is organized as follows. In Section 2 we briefly describe the sample
from which we selected the sources to be studied. In Section 3 we discuss the
method used to evaluate the thermal contribution of the selected sources. In
Section 4 we test this method on a sample of radio-quiet quasars. In Section 5
we compute the thermal spectrum for the sources in our sample and we subtract
it from the measured spectra, to disentangle the accretion disc emission from
the synchrotron one. In Section 6 we evaluate and compare the total thermal
and jet powers. In Section 7 we discuss our results while Section 8 summarizes
our conclusions. Throughout this paper the values $H_0 = 65$ Km s$^{-1}$
Mpc$^{-1}$ and $q_0 = 0.15$ have been adopted and spectral indices are written
as $f_{\nu} \propto \nu^{-\alpha}$.

\section{The Sample}

The FSRQ selected for our analysis belong to the Deep X-ray Radio Blazar
Survey (DXRBS). The DXRBS has been built by cross-correlating all the
serendipitous X-ray sources in the ROSAT database WGACAT (first revision:
White, Giommi \& Angelini 1995) with a number of radio catalogs (see Perlman
et al. 1998 and Landt et al. 2001 for details).

To select blazar candidates among all the radio sources a spectral index cut
$\alpha_r \le 0.7$ was adopted. This is a very efficient way to find blazars
because it selects all FSRQ (separated by definition from the Steep Spectrum
Radio Quasars at $\alpha_r = 0.5$), basically all BL Lacs, and excludes almost
all radio galaxies.

DXRBS is currently the faintest and largest flat-spectrum radio sample with
nearly complete ($\sim 95$ per cent) identification, reaching $\sim 50$ mJy at
5 GHz, i.e., radio fluxes $10-20$ times fainter than previously published
blazar samples. As of July 2002, the DXRBS includes 198 FSRQ; of these,
we choose the 136 sources for which the BLR luminosity, $L_{BLR}$, was
available (Landt et al., in preparation). Among these, 113 are newly
identified sources with available optical spectra. The remaining 23 FSRQ were
previously known, and for these sources we do not have optical or UV 
spectra, although we have BLR fluxes. Therefore, we calculate mass, 
accretion rate and total thermal power
for all sources (see Sect. 3 and 6), while for the former ones we also
disentangle the luminosity components in the optical band (see section 5).
    
The classification and spectra of most of the sources used in this paper 
are given in Perlman et al. (1998) and Landt et al. (2001). The remaining
sources will be included in future DXRBS papers, currently in preparation. 
Line parameters (fluxes and widths) will be presented by Landt et al., in
preparation. 

\section{the Method}\label{method}
In order to reproduce the thermal contribution to the total emission of
blazars we need an accretion disc model. These models have basically two input
parameters, the mass of the central black hole and the accretion rate.  We use
the model first proposed by Shakura \& Sunyaev (1973), which involves a
geometrically thin, optically thick steady accretion disc (Pringle 1981;
Czerny \& Elvis 1987).  The disc spectrum is the following:

$$L_{{\nu}, th} = \pi \int_{fR_{G}}^{R_{out}} {2\,h\,\nu^3\over c^2}\;
{2\,\pi\,R\,dR\over e^{h\nu/KT_S(R)}-1},\eqno(1)$$ 
where $R_G = GM/c^2$ is the gravitational radius, $f\,R_G$ is the radius of
the innermost stable orbit and $R_{out}$ is the maximum distance from the
central black hole at which the spectrum is computed.  Finally, $T_S(R)$ is
the surface temperature of the accretion disc and is given by:

$$T_S(R) = \left[ {3\,G\,M\,\dot M \over 8\,\pi\,R^3\,\sigma} \left(1 - \left( R_G\over R\right)^{1/2}\right)\right]^{1/4},\eqno(2)$$
where $\sigma$ is the Stefan constant and $M$ and $\dot M$ are the mass of the
central black hole and the accretion rate, respectively.

In principle, one can evaluate $M$ using the formula

$$M = {V^2\,R\over aG}, \eqno (3)$$ 
once the distance $R$ and the velocity $V$ of a body in the gravitational
field of the black hole are known. Here $a=2$ for a body in free-fall and
$a=1$ for an orbiting body. While the velocity of the gas in the BLR can be
evaluated quite easily from observed spectra using the line widths, the
determination of the distance $R$ is not straightforward. Basically, two
different methods can be used (see Wandel, Peterson \& Malkan 1999). The first
one is indirect and relies on photoionization theory (e.g., Netzer 1990) by
using the definition of the ionization parameter $U$ (proportional to the flux
of ionizing photons per electron):

$$U = {Q\over4\,\pi\,R^2\,n_e\,c},\eqno (4)$$  
where 

$$Q=\int_{13.6\,eV}^{\infty}{L_{\nu}\over h\nu}d\nu\eqno(5)$$ 
is the number of ionizing photons per unit of time and $n_e$ is the electronic
density in the shell of the BLR at distance $R$ from the center.  The second
method, more direct, takes advantage of the reverberation-mapping technique
(e.g., Peterson 1993).  This allows one to evaluate the radius of the BLR by
measuring the time lags between variations in the broad emission lines and the
ionizing continuum.

We can see that in order to evaluate $R$ from eq. (4) we need to know $U\,n_e$
and $Q$.  The latter has been calculated as follows. We assume for our sources
an average optical to X-ray effective spectral index $\alpha_{\rm ox} = 1.56$,
equal to that measured by Laor et al. (1997) for the 19 radio-quiet quasars
from the Bright Quasar Survey. We then substitute in eq. (5) $L_{\nu} =
L_{13.6}(\nu/\nu_{13.6})^{-\alpha_{\rm ox}}$ and calculate $L_{13.6}$ for
every single object using the equation:

$$L_{ion} = f_{\rm cov}^{-1}\, L_{BLR} = \int_{13.6\,eV}^{\infty} L_{\nu}d\nu, \eqno (6)$$
where $L_{ion}$ is the power of the ionizing continuum and $f_{\rm cov}$ is
the BLR covering factor, which we assume to be $\sim 10$ per cent (see next
subsection).

The second quantity needed, $U\,n_e$, varies from line to line, due to the
complex structure of the BLR. This parameter can be estimated from eq. (4) if
the distance of the clouds emitting a particular line is known. This is
available for a sizeable sample of source for H$\beta$ (e.g., Wandel et
al. 1999 and references therein) since this line is easy to observe in
low-redshift, and better studied, AGN. By comparing the radius measured with
the reverberation mapping techniques with that evaluated with the
photoionization model for 18 sources we derive $(U\,n_e)_{H\beta} \sim 5
\times 10^{9}$ cm$^{-3}$. However, our sample has an average redshift $z \sim
1.5$, which means that only a relatively small number of sources (32) has
$H\beta$ information. We then need an estimate of $U\,n_e$ for some lines in
the UV band, which unfortunately have not been studied exhaustively
yet. Fromerth and Melia (2000) have recently published the distances of the C
IV line region for five AGN, evaluated using the reverberation mapping
technique. We can then use these values and the H$\beta$ distances for the
same objects (Wandel et al. 1999) to evaluate $(U\,n_e)_{CIV}$ as follows:

$$ (U\,n_e)_{CIV} = (U\,n_e)_{H\beta}\,\left({R_{H\beta}\over R_{CIV}}\right)^2. \eqno (7)$$

We perform this calculation for all the five objects and we then compute an
average to obtain $(U\,n_e)_{CIV} = 1.34 \times (U\,n_e)_{H\beta} \sim 6.7
\times 10^{9}$ cm$^{-3}$. With this value we can evaluate $R_{CIV}$ for 40
more objects (the ones showing the C IV line).

\begin{figure}%1
\epsfysize=9cm 
\hspace{3.5cm}\epsfbox{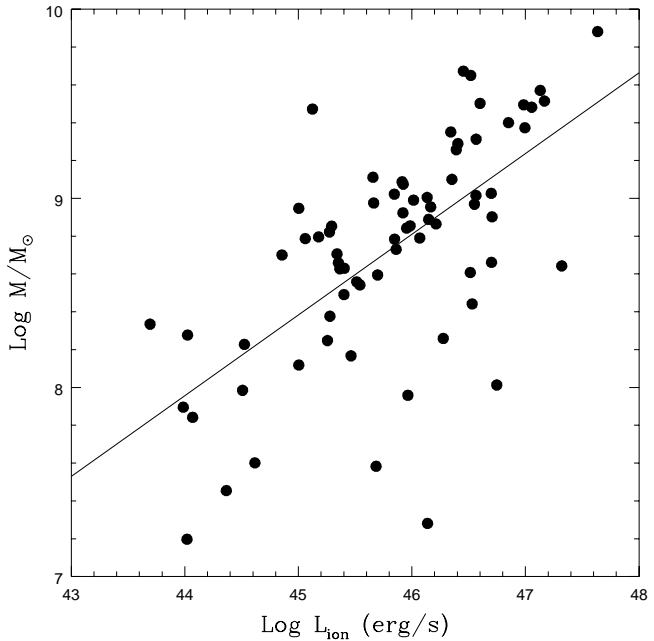}
\caption[h]{The mass--ionizing power relation for the 72 FSRQ having H$\beta$
or C IV line in their spectra. The solid line is the best fit to the data:
$\log (M/M_{\odot}) = 0.43 \,\log L_{ion} - 10.8$.\label{mass_lion}}
\end{figure}

{}From the full width half maximum (FWHM) values and from eq. (3) we calculate
the masses for these 72 objects, using $a = 1$ and $V = 1.5$ FWHM, which
represents a reasonable choice for radio-loud quasars, as suggested by McLure
and Dunlop (2001). In a handful of cases (and only for H$\beta$) we found a
relatively strong narrow component associated with the line which, if not
subtracted, would have given a spuriously low value for the FWHM (and
therefore the mass). In those cases the FWHM values we are using refer to the
broad component of the lines. The relation between the derived mass and the
ionizing power L$_{ion}$ is plotted in Figure \ref{mass_lion}. We find a very
strong correlation ($P \sim 99.99 $ per cent) between the two quantities, with
$M \propto L_{ion}^{0.43\pm0.06}$, and then use this correlation to evaluate 
the
masses for the 64 FSRQ without H$\beta$ or C IV information. Note that such a
dependence is not unexpected: in our sample we find no correlation between
FWHM and L$_{ion}$ and therefore, since we assume $R_{BLR} \propto Q^{0.5}
\propto L^{0.5}$, we will obtain $M \proptosim L_{ion}^{0.5}$.

The second parameter needed to compute any accretion disc model is the
accretion rate of mass onto the central black hole ($\dot M$). This quantity
is calculated using the following relation:

$$ L_{ion} = \epsilon \, \dot M \,c^2, \eqno (8)$$
where $\epsilon$ is the efficiency parameter. In the following we assume that
the black holes in the center of our FSRQ are rapidly rotating, so that we
choose $\epsilon \sim 0.3$ (Malkan 1982).

We finally evaluate the Eddington ratio for all the 136 sources for which mass
and accretion rate have been calculated:

$${L\over L_{Edd}} = 4.5\,\epsilon \, {\dot M \over M_8}, \eqno (9)$$
where $\dot M$ is expressed in $M_{\odot}/yr$ and $M_8 = 10^8\, M_{\odot}$.
Table 1 gives masses, accretion rates, and Eddington ratios for all our FSRQ.
Note that masses span the range $2 \times 10^7 - 7 \times 10^{9}\, M_{\odot}$,
with $\langle \log M \rangle = 8.74\pm0.04\, M_{\odot}$, while accretion rates
cover the range $3 \times 10^{-3} - 72\, M_{\odot}/yr$, with $\langle \log
\dot M \rangle = -0.40\pm0.07 M_{\odot}/yr$.

\begin{table*}
\scriptsize
\begin{minipage}{180mm}
\centerline{\bf Table 1. Masses, Accretion Rates, Eddington Ratios and Luminosities}
\hspace{1.0cm} 
\hspace{1.0cm} 
\begin{tabular}{|l|r|r|r|r|c|r|c|r|c|}
\hline
Name& $\log M$  &$\log \dot M$ &$\log L/L_{Edd}$ &$\log L_{th}/L_{tot}^{(1)}$&$\log L_{tot}^{(1)}$ 
			&$\log L_{th}/L_{tot}^{(2)}$&$\log L_{tot}^{(2)}$&$\log L_{th}/L_{tot}^{(3)}$&$\log L_{tot}^{(3)}$  \\
         & $M_{\odot}$ & $M_{\odot}/yr$ & & & erg s$^{-1}$ Hz$^{-1}$ & &  
erg s$^{-1}$ Hz$^{-1}$ & &  erg s$^{-1}$ Hz$^{-1}$ \\

\hline
WGAJ0010.5$-$3027$^c$    & 8.8&-0.3& -0.9& ...& ...&-1.0&30.89& ...& ...\\
WGAJ0010.7$-$3649$^b$    & 9.3& 0.2& -1.0& ...& ...&-1.6&32.04&-1.4&31.78\\
WGAJ0011.2$-$3620$^b$    & 8.8&-0.3& -1.0& ...& ...& ...& ...&-0.4&30.28\\
WGAJ0012.5$-$1629$^a$    & 7.2&-2.2& -1.3&-1.5&28.98& ...& ...& ...& ...\\
WGAJ0014.5$-$3059$^b$    & 9.5& 0.4& -1.0& ...& ...& ...& ...&-1.0&31.66\\
WGAJ0029.0+0509$^b$      & 9.4& 0.1& -1.1& ...& ...&-0.9&31.31& ...& ...\\
WGAJ0032.5$-$2648$^c$    & 8.9& 0.1& -0.8& ...& ...&-0.4&30.58& ...& ...\\
WGAJ0049.5$-$2509$^b$    & 7.5&-1.9& -1.2& ...& ...& ...& ...& ...& ...\\
PKS 0100$-$270$^b$       & 8.9&-0.2& -1.0& ...& ...& ...& ...& ...& ...\\
WGAJ0106.7$-$1034$^a$    & 8.7&-0.9& -1.4&-0.7&30.02& ...& ...& ...& ...\\
WGAJ0110.5$-$1647$^c$    & 8.9&-0.1& -0.8&-0.8&30.82&-0.8&30.88& ...& ...\\
1Jy0112$-$017$^b$        & 9.0&-0.6& -1.4& ...& ...& ...& ...& ...& ...\\
1Jy0119+041$^c$          & 8.6&-0.6& -1.1& ...& ...& ...& ...& ...& ...\\
UM 320$^b$               & 8.9&-0.1& -0.9& ...& ...& ...& ...& ...& ...\\
WGAJ0126.2$-$0500$^a$    & 8.2&-1.7& -1.8&-0.9&29.34& ...& ...& ...& ...\\
WGAJ0136.0$-$4044$^c$    & 8.3&-1.5& -1.6&-1.2&29.79& ...& ...& ...& ...\\
WGAJ0143.2$-$6813$^c$    & 8.6&-0.8& -1.3& ...& ...&-0.7&30.02& ...& ...\\
WGAJ0146.2+0223$^b$      & 8.3& 0.0& -0.1& ...& ...&-1.6&31.39&-1.5&31.28\\
WGAJ0210.0$-$1004$^b$    & 9.0&-0.4& -1.3& ...& ...&-1.0&30.93&-0.9&30.75\\
WGAJ0210.1$-$3851$^b$    & 9.5& 0.8& -0.5& ...& ...&-0.6&31.68& ...& ...\\
WGAJ0216.6$-$7331$^b$    & 9.0& 0.5& -0.4& ...& ...& ...& ...&-0.8&31.30\\
WGAJ0217.7$-$7347$^c$    & 8.7&-0.4& -1.0& ...& ...&-0.7&30.43& ...& ...\\
WGAJ0227.5$-$0847$^b$    & 9.6& 0.9& -0.6& ...& ...&-0.5&31.67&-0.4&31.52\\
PKS 0256$-$005$^c$       & 8.9&-1.2& -2.1&-0.3&29.55& ...& ...& ...& ...\\
WGAJ0258.6$-$5052$^c$    & 9.2& 0.8& -0.4& ...& ...& ...& ...& ...& ...\\
WGAJ0259.4+1926$^a$      & 8.2&-1.7& -1.7&-1.9&30.35& ...& ...& ...& ...\\
WGAJ0304.9+0002$^a$      & 8.5&-0.8& -1.2&-0.7&29.92&-0.6&29.92& ...& ...\\
WGAJ0312.3$-$6610$^c$    & 8.5&-1.0& -1.4& ...& ...&-1.0&30.18& ...& ...\\
WGAJ0314.4$-$6548$^a$    & 8.7&-0.9& -1.5&-0.5&29.84& ...& ...& ...& ...\\
WGAJ0322.1$-$5205$^a$    & 8.9&-0.9& -1.7& ...& ...&-0.7&30.11& ...& ...\\
WGAJ0322.2$-$5042$^a$    & 8.6&-0.9& -1.4&-0.5&29.81& ...& ...& ...& ...\\
WGAJ0322.6$-$1335$^b$    & 8.8&-1.0& -1.7&-0.3&29.61&-0.3&29.70& ...& ...\\
WGAJ0324.9$-$2140$^b$    & 9.9& 1.4& -0.4& ...& ...&-0.4&32.07&-0.3&32.01\\
WGAJ0357.6$-$4158$^c$    & 8.7&-0.5& -1.1& ...& ...&-0.6&30.24& ...& ...\\
WGAJ0411.0$-$1637$^a$    & 9.1& 0.1& -0.9& 0.0&30.31& ...& ...& ...& ...\\
WGAJ0424.6$-$3849$^b$    & 8.0&-0.3& -0.1& ...& ...&-0.6&29.98&-0.4&29.82\\
WGAJ0427.2$-$0756$^b$    & 9.7& 0.2& -1.3& ...& ...& 0.2&30.50& ...& ...\\
WGAJ0434.3$-$1443$^b$    & 9.1&-0.3& -1.3& ...& ...&-1.5&31.45&-1.0&30.93\\
WGAJ0435.1$-$0811$^a$    & 7.6&-0.5&  0.0&-1.0&29.83&-1.0&29.92& ...& ...\\
WGAJ0441.8$-$4306$^c$    & 8.1&-1.8& -1.8&-1.2&29.48&-0.6&28.98& ...& ...\\
WGAJ0447.9$-$0322$^c$    & 9.2& 0.6& -0.5&-0.7&31.31&-0.7&31.38& ...& ...\\
WGAJ0448.2$-$2110$^b$    & 8.7& 0.5& -0.1& ...& ...&-1.2&31.50&-0.8&31.11\\
WGAJ0448.6$-$2203$^c$    & 8.0&-2.2& -2.1&-0.5&28.46& ...& ...& ...& ...\\
WGAJ0510.0+1800$^c$      & 8.2&-1.6& -1.7&-1.8&30.29& ...& ...& ...& ...\\
1Jy0514$-$459$^a$        & 7.6&-1.6& -1.1& ...& ...& ...& ...& ...& ...\\
WGAJ0535.1$-$0239$^c$    & 8.9& 0.1& -0.8&-0.5&30.67&-0.4&30.60& ...& ...\\
WGAJ0539.0$-$3427$^a$    & 8.3&-2.5& -2.8&-0.4&28.32& ...& ...& ...& ...\\
WGAJ0546.6$-$6415$^a$    & 9.1&-0.3& -1.3&-0.5&30.50& ...& ...& ...& ...\\
WGAJ0600.5$-$3937$^c$    & 8.9& 0.0& -0.8& ...& ...&-1.0&31.14&-0.9&31.11\\
WGAJ0631.9$-$5404$^a$    & 9.5&-1.1& -2.5&-0.2&29.74& ...& ...& ...& ...\\
WGAJ0648.2$-$4347$^c$    & 9.0& 0.1& -0.7&-0.9&31.06&-0.7&30.94& ...& ...\\
S5 0743+74$^c$           & 8.9& 0.0& -0.8&-1.6&31.71&-1.3&31.47& ...& ...\\
WGAJ0744.8+2920$^c$      & 8.8&-0.2& -0.9& ...& ...& ...& ...& ...& ...\\
WGAJ0747.0$-$6744$^c$    & 8.6&-0.8& -1.3& ...& ...&-0.4&29.74& ...& ...\\
WGAJ0748.2$-$5257$^b$    & 9.3& 0.3& -0.9& ...& ...&-0.8&31.36&-0.7&31.30\\
WGAJ0751.0$-$6726$^c$    & 8.9& 0.0& -0.8&-0.7&30.77&-0.7&30.86& ...& ...\\
1Jy0850+581$^c$          & 9.3& 0.8& -0.3& ...& ...& ...& ...& ...& ...\\
OJ$-$297$^c$             & 9.1& 0.5& -0.5& ...& ...& ...& ...& ...& ...\\
1Jy0859+470$^b$          & 8.0& 0.5&  0.6& ...& ...& ...& ...& ...& ...\\
WGAJ0927.7$-$0900$^a$    & 7.8&-2.2& -1.9&-1.1&28.96& ...& ...& ...& ...\\
WGAJ0937.2+5008$^a$      & 8.3&-2.2& -2.4&-1.5&29.65& ...& ...& ...& ...\\
WGAJ1003.9+3244$^b$      & 9.3& 0.2& -1.0& ...& ...&-0.3&30.73& ...& ...\\
WGAJ1006.5+0509$^c$      & 8.5&-0.9& -1.3& ...& ...& 0.1&29.16& ...& ...\\
WGAJ1010.8$-$0201$^a$    & 8.8&-0.4& -1.0&-0.7&30.40&-0.5&30.27& ...& ...\\
WGAJ1011.5$-$0423$^c$    & 9.0& 0.1& -0.7& ...& ...&-0.9&31.08&-0.7&30.96\\
WGAJ1025.9+1253$^a$      & 9.3& 0.8& -0.3& 0.4&30.42& ...& ...& ...& ...\\
WGAJ1026.4+6746$^c$      & 8.2&-1.0& -1.1&-1.0&29.97& ...& ...& ...& ...\\
S5 1027+74$^a$           & 8.1&-1.9& -1.9&-1.8&30.07&-1.1&29.41& ...& ...\\
WGAJ1028.5$-$0236$^c$    & 7.9&-2.2& -2.0& ...& ...& ...& ...& ...& ...\\
WGAJ1032.1$-$1400$^c$    & 8.7&-0.4& -1.0&-0.9&30.56&-0.7&30.41& ...& ...\\
WGAJ1035.0+5652$^c$      & 8.3&-1.4& -1.6&-0.9&29.59& ...& ...& ...& ...\\
WGAJ1046.3+5354$^b$      & 9.0& 0.3& -0.5& ...& ...&-0.5&30.92& ...& ...\\
B2 1048+34$^c$           & 9.1& 0.5& -0.5& ...& ...& ...& ...& ...& ...\\
WGAJ1101.8+6241$^a$      & 8.2&-0.8& -0.8&-0.8&29.82& ...& ...& ...& ...\\
WGAJ1104.8+6038$^c$      & 9.0& 0.3& -0.6& ...& ...&-0.4&30.75& ...& ...\\
WGAJ1105.3$-$1813$^a$    & 8.4&-1.0& -1.2&-0.7&29.73&-0.6&29.72& ...& ...\\
WGAJ1112.5$-$3745$^c$    & 8.8&-0.2& -0.9&-0.8&30.67&-0.6&30.51& ...& ...\\
\hline
\end{tabular}

\end{minipage}  
\end{table*}

\begin{table*}
\scriptsize
\begin{minipage}{180mm}
\centerline{\bf Table 1. (Continued)}
\hspace{1.0cm} 
\begin{tabular}{|l|r|r|r|r|c|r|c|r|c|}
\hline
Name& $\log M$  &$\log \dot M$ &$\log L/L_{Edd}$ &$\log L_{th}/L_{tot}^{(1)}$&$\log L_{tot}^{(1)}$ 
			&$\log L_{th}/L_{tot}^{(2)}$&$\log L_{tot}^{(2)}$&$\log L_{th}/L_{tot}^{(3)}$&$\log L_{tot}^{(3)}$  \\
         & $M_{\odot}$ & $M_{\odot}/yr$ & & & erg s$^{-1}$ Hz$^{-1}$ & &  
erg s$^{-1}$ Hz$^{-1}$ & &  erg s$^{-1}$ Hz$^{-1}$ \\

\hline
WGAJ1116.1+0828$^c$      & 8.0&-2.0& -2.0&-1.8&29.89& ...& ...& ...& ...\\
WGAJ1150.4+0156$^c$      & 9.0& 0.2& -0.7& ...& ...&-0.2&30.57& ...& ...\\
WGAJ1206.2+2823$^c$      & 8.6&-0.8& -1.3&-0.6&29.87&-0.4&29.70& ...& ...\\
WGAJ1213.2+1443$^c$      & 8.7&-0.6& -1.1&-0.3&29.75& ...& ...& ...& ...\\
WGAJ1217.1+2925$^c$      & 8.5&-1.0& -1.3& ...& ...&-0.3&29.47& ...& ...\\
WGAJ1223.9+0650$^c$      & 8.6&-0.6& -1.2& ...& ...&-0.6&30.15& ...& ...\\
WGAJ1300.7$-$3253$^c$    & 8.8&-0.2& -0.9& ...& ...&-1.0&30.45& ...& ...\\
WGAJ1314.0$-$3304$^c$    & 8.5&-0.9& -1.3&-0.5&29.70& ...& ...& ...& ...\\
WGAJ1315.1+2841$^c$      & 8.9& 0.0& -0.8& ...& ...&-0.9&30.97& ...& ...\\
WGAJ1324.0$-$3623$^a$    & 8.8&-0.2& -0.8&-0.9&30.77&-0.7&30.67& ...& ...\\
WGAJ1333.1$-$3323$^b$    & 9.0&-0.2& -1.1& ...& ...&-0.6&30.57&-0.5&30.49\\
WGAJ1337.2$-$1319$^b$    & 9.4& 0.8& -0.5& ...& ...&-1.1&32.07&-1.0&31.99\\
WGAJ1359.6+4010$^a$      & 8.0&-1.7& -1.6&-1.7&30.03& ...& ...& ...& ...\\
WGAJ1400.7+0425$^b$      & 8.6&-0.5& -1.0& ...& ...& ...& ...&-1.2&30.83\\
WGAJ1402.7$-$3334$^b$    & 8.8&-1.1& -1.7& ...& ...&-1.7&30.99&-1.3&30.58\\
WGAJ1406.9+3433$^b$      & 9.5& 0.9& -0.5& ...& ...&-0.8&31.95&-0.5&31.67\\
WGAJ1416.4+1242$^a$      & 8.8&-1.2& -1.8&-0.4&29.53& ...& ...& ...& ...\\
WGAJ1417.5+2645$^b$      & 8.6&-0.8& -1.3& ...& ...&-0.7&30.06& ...& ...\\
WGAJ1419.1+0603$^b$      & 9.0&-0.1& -0.9& ...& ...&-0.6&30.73&-0.6&30.73\\
WGAJ1442.3+5236$^c$      & 9.0& 0.3& -0.6& ...& ...&-0.8&31.27& ...& ...\\
WGAJ1457.7$-$2818$^b$    & 9.7& 0.3& -1.2& ...& ...&-0.5&31.22&-0.5&31.14\\
1Jy1502+106$^c$          & 9.0& 0.3& -0.6& ...& ...& ...& ...& ...& ...\\
WGAJ1506.6$-$4008$^c$    & 8.9& 0.0& -0.8&-0.7&30.73&-0.6&30.73& ...& ...\\
WGAJ1509.5$-$4340$^a$    & 8.7&-0.4& -1.0&-0.6&30.28&-0.7&30.41& ...& ...\\
WGAJ1525.3+4201$^c$      & 8.9& 0.0& -0.8& ...& ...&-0.9&30.96& ...& ...\\
WGAJ1543.6+1847$^c$      & 8.8&-0.2& -0.9& ...& ...&-0.5&30.46& ...& ...\\
WGAJ1610.3$-$3958$^a$    & 8.1&-1.2& -1.2&-1.7&30.46& ...& ...& ...& ...\\
OS319$^b$                & 9.4& 0.6& -0.7& ...& ...& ...& ...& ...& ...\\
WGAJ1615.2$-$0540$^b$    & 8.9&-0.3& -1.1& ...& ...& ...& ...&-2.6&32.55\\
WGAJ1629.7+2117$^c$      & 8.4&-1.1& -1.4& ...& ...& ...& ...& ...& ...\\
4C38.41$^b$              & 8.6& 1.1&  0.6& ...& ...& ...& ...& ...& ...\\
1Jy1637+574$^a$          & 8.4& 0.3&  0.0& ...& ...& ...& ...& ...& ...\\
1Jy1638+398$^c$          & 8.5&-0.9& -1.3& ...& ...& ...& ...& ...& ...\\
3C345$^a$                & 8.9& 0.5& -0.3& ...& ...& ...& ...& ...& ...\\
WGAJ1656.6+5321$^c$      & 8.9&-0.1& -0.8& ...& ...&-0.3&30.33& ...& ...\\
WGAJ1656.6+6012$^a$      & 8.5&-0.7& -1.1&-0.7&30.07& ...& ...& ...& ...\\
1Jy1725+044$^c$          & 8.6&-0.8& -1.2& ...& ...& ...& ...& ...& ...\\
WGAJ1738.6$-$5333$^b$    & 9.5& 0.8& -0.6& ...& ...&-1.1&32.07&-0.8&31.76\\
WGAJ1804.7+1755$^a$      & 8.6&-0.7& -1.2&-0.7&30.08& ...& ...& ...& ...\\
WGAJ1808.2$-$5011$^b$    & 8.9& 0.0& -0.8& ...& ...&-0.8&30.86& ...& ...\\
WGAJ1826.1$-$3650$^c$    & 8.6&-0.8& -1.3&-1.4&30.70&-1.1&30.40& ...& ...\\
WGAJ1827.1$-$4533$^c$    & 9.0& 0.2& -0.7& ...& ...&-0.7&31.04& ...& ...\\
WGAJ1837.7$-$5848$^b$    & 9.0& 0.3& -0.6& ...& ...& ...& ...&-1.0&31.45\\
WGAJ1911.8$-$2102$^c$    & 8.8&-0.3& -1.0& ...& ...&-1.0&30.76& ...& ...\\
PKS 1937$-$101$^c$       & 9.7& 1.9&  0.3& ...& ...& ...& ...& ...& ...\\
PKS 2059+034$^c$         & 8.9& 0.0& -0.8& ...& ...& ...& ...& ...& ...\\
WGAJ2109.7$-$1332$^c$    & 9.1& 0.4& -0.5&-0.8&31.32&-0.7&31.21& ...& ...\\
WGAJ2154.1$-$1502$^c$    & 9.2& 0.8& -0.4&-0.7&31.47&-0.6&31.41& ...& ...\\
WGAJ2157.7+0650$^a$      & 8.7&-1.4& -2.0&-0.7&29.67& 0.0&29.02& ...& ...\\
WGAJ2159.3$-$1500$^b$    & 8.6& 0.3& -0.2& ...& ...&-1.5&31.65&-1.3&31.51\\
OY$-$106$^c$             & 8.4&-1.1& -1.4& ...& ...& ...& ...& ...& ...\\
WGAJ2248.6$-$2702$^b$    & 9.1&-0.6& -1.6& ...& ...&-1.9&31.73&-1.4&31.18\\
WGAJ2320.6+0032$^b$      & 9.0&-0.1& -1.0& ...& ...&-1.5&31.56&-0.9&30.97\\
WGAJ2322.0+2113$^a$      & 7.3&-0.1&  0.7&-0.6&30.55&-0.6&30.66& ...& ...\\
WGAJ2329.0+0834$^c$      & 8.4&-1.2& -1.5& ...& ...&-0.8&29.69& ...& ...\\
WGAJ2333.2$-$0131$^c$    & 8.7&-0.6& -1.1& ...& ...&-0.3&29.86& ...& ...\\
1Jy2344+092$^c$          & 9.1& 0.5& -0.5& ...& ...& ...& ...& ...& ...\\
WGAJ2349.9$-$2552$^c$    & 8.7&-0.5& -1.1&-0.9&30.40&-0.5&30.10& ...& ...\\
WGAJ2354.2$-$0957$^c$    & 8.3&-1.5& -1.7&-1.5&30.13& ...& ...& ...& ...\\
\hline

\end{tabular}

$^{(1)}$ Values calculated in the interval $7.08 \times 10^{14} - 7.24 \times
10^{14}$ Hz.
 
$^{(2)}$ Values calculated in the interval $1.32 \times 10^{15} - 1.35 \times
10^{15}$ Hz. 

$^{(3)}$ Values calculated in the interval $2.24 \times 10^{15} - 2.29 \times
10^{15}$ Hz.

$^a$ Sources whose mass has been calculated using the H$\beta$ line.

$^b$ Sources whose mass has been calculated using the C IV line.

$^c$ Sources whose mass has been calculated using the correlation luminosity-mass (see Fig. 
\ref{mass_lion}).
\end{minipage}  
\end{table*}

\begin{figure}%2
\epsfysize=9cm 
\hspace{3.5cm}\epsfbox{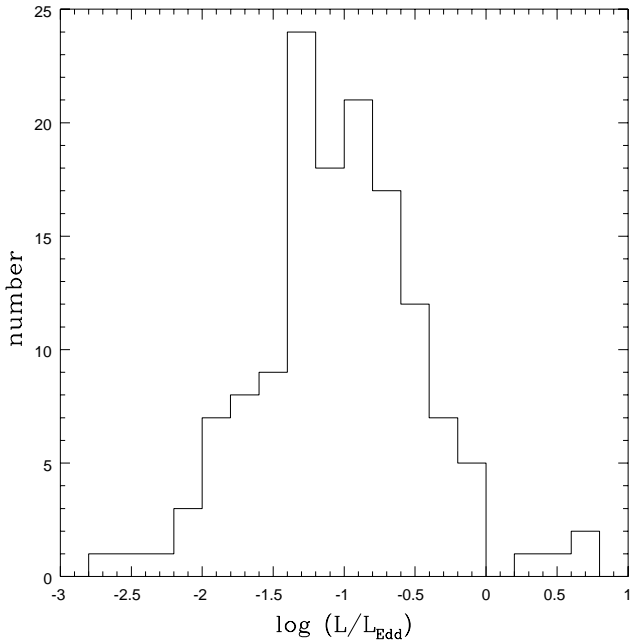}
\caption[h]{The distribution of the Eddington ratio $L/L_{Edd}$ for our 
FSRQ.\label{edd_distrib}}
\end{figure}

Figure \ref{edd_distrib} shows the distribution of the Eddington ratio, which
lies between $0.002$ and $1$ for almost all the FSRQ, with $\langle \log
L/L_{Edd}\rangle = -1.02\pm0.05$. Only four sources ($\sim 3\%$) have
$L/L_{Edd} > 1$, which is reassuring. The average value for the 32 H$\beta$
sources is $10^{-1.3\pm0.1}$, that for the 40 C IV ones is $10^{-0.81\pm0.09}$
and the one for the remaining 64 sources is $10^{-1.0\pm0.1}$.

Note that by using eq. (4) we assume that $R_{BLR} \propto Q^{0.5} \propto
L^{0.5}$. This is in apparent contrast with Kaspi et al. (2000), who found
$R_{BLR} \propto L^{0.70\pm0.03}$. However, Vestergaard (2002), by properly
taking into account measurement uncertainties in both $R_{BLR}$ and $L$, has
obtained a flatter dependency with a slope of $0.58\pm0.09$, fully consistent
with our assumption. 

Recently, Vestergaard (2002) has also derived a relationship to estimate AGN
masses based on UV luminosity and the FWHM of C IV. This was calibrated on
recent estimates of low-redshift AGN masses obtained via reverberation mapping
using the $H\beta$ line. A comparison between our mass estimates for the 40
objects with C IV data in our sample and those derived using Vestergaard's
relationship shows that the two are in extremely good agreement, with a
basically linear dependence ($1.16\pm0.04$). The means for the two methods are
$8.99\pm0.08$ and $8.98\pm0.09$ respectively, and the two distributions are
not significantly different according to a Kolmogorov-Smirnov (KS) test.

\subsection{Covering Factor}

The covering factor of the BLR in (mostly radio-quiet) AGN has been discussed
by several authors in the past (see Maiolino et al. 2001 and references
therein). Early estimates of this quantity seemed to indicate a covering
factor $f_{\rm cov} \sim 5 - 10$ per cent, while recent observations indicate
instead $ f_{\rm cov} \sim 30$ per cent. Estimates of the covering factor need
measurements of the UV continuum emission from the accretion disc.  In the
case of blazars this emission is swamped by the jet and this makes the
covering factor a difficult quantity to calculate for these sources. The only
way to get a rough estimate of $ f_{\rm cov}$ is to know the luminosity of the
BLR of objects in which the thermal emission is clearly dominating the
synchrotron one. We find in literature only two of these objects: 3C 273 and
PKS 2149$-$306. For the first source (see Kriss et al. 1999 and Celotti et
al. 1997 for the data) we estimate $ f_{\rm cov} \sim 7$ per cent, while for
the second one (Elvis et al. 2000; Wilkes et al. 1983) we find $ f_{\rm cov}
\sim 10$ per cent. For this reason, we adopt $ f_{\rm cov} \sim 10$ per cent
for our sources and we comment on how our results would change with $ f_{\rm
cov} = 30$ per cent.

\section{Testing the Method}
Before computing the thermal contribution for our FSRQ sample, we need to know
if our method is reliable. The best way to do this is to evaluate masses and
accretion rates for a sample of radio-quiet quasars (which are not affected by
beaming), compute their thermal emission, and compare it with the data. All we
need to know for such a test is the power of the BLR, the effective spectral
index $\alpha_{\rm ox}$, the FWHM of H$\beta$ or C IV and a measure of the
luminosity in the optical-UV band. Laor et al. (1997) give for 19 low-redshift
radio-quiet quasars the luminosity at $10^{15}$ Hz, $\alpha_{\rm ox}$, and the
FWHM and power of H$\beta$ (see their Table 4; we excluded the 4
radio-loud quasars in their list). The only missing parameter is $L_{BLR}$,
but we can estimate it assuming $L_{H\beta}\, / \,L_{BLR} \sim 0.04$ (Francis
et al. 1991; Celotti et al. 1997).

We use eq. (3) to evaluate the masses and eq. (8) with $\epsilon = 0.057$
(assuming non-rotating black holes, see Misner, Thorne \& Wheeler 1970) to
evaluate the accretion rates; we then compute the accretion disc spectra using
eq. (1) with $f = 6$, because in non-rotating black holes (which is likely the
case in radio-quiet sources; e.g., Wilson \& Colbert 1995; Moderski, Sikora \&
Lasota 1998) the main stable orbit is roughly located at six times the
gravitational radius.

\begin{table*}
\begin{minipage}{180mm}
\centerline{\bf Table 2. Masses, Accretion Rates and Luminosities for Bright Quasar Survey Sources}
\hspace{1.0cm} 
\hspace{1.0cm} 
\begin{tabular}{|l|c|c|c|c|c|c|c|}
\hline
Name (PG)& $M_8$  &$\dot M$ &$\nu_P$&$L_{th}(\nu_P)$  
			&$L_{ext}(\nu_P)$&$\log (L_{ext}/L_{th})$\\
         &        & $(M_{\odot}/yr)$ & (Hz) & (erg s$^{-1}$ Hz$^-1$) & 
(erg s$^{-1}$ Hz$^-1$) & \\

\hline
0947+396      &$14$&$1.8  $&$1.2\,10^{15}$&$2.0\,10^{30}$&$1.3\,10^{30}$&$-0.200$\\
0953+414      &$14$&$10  $&$1.8\,10^{15}$&$7.3\,10^{30}$&$3.1\,10^{30}$&$-0.372$\\
1001+054       &$1.0$&$1.4  $&$4.2\,10^{15}$&$4.4\,10^{29}$&$5.4\,10^{28}$&$-0.911$\\
1048+342        &$11$&$3.3  $&$1.6\,10^{15}$&$2.7\,10^{30}$&$6.9\,10^{29}$&$-0.598$\\   
1114+445     &$12$&$1.6  $&$1.3\,10^{15}$&$1.6\,10^{30}$&$8.2\,10^{29}$&$-0.294$\\
1115+407    &$1.1$&$0.66  $&$3.3\,10^{15}$&$2.6\,10^{29}$&$1.5\,10^{29}$&$-0.233$\\
1116+215 &$13$&$11  $&$2.0\,10^{15}$&$7.3\,10^{30}$&$2.5\,10^{30}$&$-0.457$\\
1202+281 &$14$&$1.5  $&$1.1\,10^{15}$&$1.7\,10^{30}$&$8.9\,10^{29}$&$-0.286$\\
1216+069 &$39$&$10  $&$1.1\,10^{15}$&$1.2\,10^{31}$&$7.6\,10^{30}$&$-0.205$\\
1322+659 &$4.0$&$1.3  $&$2.1\,10^{15}$&$8.1\,10^{29}$&$7.2\,10^{29}$&$-0.051$\\
1352+183 &$9.0$&$2.8  $&$1.7\,10^{15}$&$2.2\,10^{30}$&$8.4\,10^{29}$&$-0.417$\\
1402+261 &$2.2$&$1.8  $&$3.0\,10^{15}$&$7.8\,10^{29}$&$5.4\,10^{29}$&$-0.164$\\
1411+442 &$3.6$&$1.2  $&$2.2\,10^{15}$&$7.2\,10^{29}$&$1.6\,10^{29}$&$-0.645$\\
1415+451 &$2.2$&$0.51  $&$2.2\,10^{15}$&$3.0\,10^{29}$&$1.9\,10^{29}$&$-0.201$\\
1427+480 &$4.3$&$2.1  $&$2.5\,10^{15}$&$1.6\,10^{30}$&$4.4\,10^{29}$&$-0.552$\\
1440+356 &$0.77$&$0.66  $&$4.0\,10^{15}$&$2.1\,10^{29}$&$1.1\,10^{29}$&$-0.310$\\
1444+407 &$4.6$&$2.5  $&$2.3\,10^{15}$&$1.4\,10^{30}$&$1.1\,10^{30}$&$-0.101$\\
1543+489 &$2.4$&$4.8  $&$3.7\,10^{15}$&$1.7\,10^{30}$&$6.9\,10^{29}$&$-0.383$\\
1626+554 &$11$&$1.5  $&$1.3\,10^{15}$&$1.5\,10^{30}$&$7.3\,10^{29}$&$-0.320$\\
\hline

\end{tabular}

\end{minipage}  
\end{table*}

To perform a realistic comparison between the data and our model we take into
account the fact that in radio-quiet quasar spectra the thermal contribution
emerges from a power law (see Malkan 1982; Band \& Malkan 1989) and thus it is
dominant only around its peak. For this reason, we extrapolate the observed
luminosity at the frequency of the maximum of the computed thermal emission,
using for each source its $\alpha_{\rm ox}$.

In Table 2 we give, for each of the 19 sources, the computed thermal
luminosity, the extrapolated luminosity from the data at the peak frequency
($\nu_P$) of the thermal emission and the ratio between the two powers.

The average logarithmic ratio is $-0.35\pm 0.05$; this means that the observed
luminosity is, on average, $\sim 50$ per cent smaller than that computed with
the accretion disc model. If we use $f_{\rm cov} = 30$ per cent, which should
represent the optimal choice for radio-quiet AGN as discussed in Section 3.1,
the average logarithmic ratio modifies to $0.13 \pm 0.05$. That is, the
observed luminosity is, on average, $\sim 35$ per cent larger than that from
the accretion disc model. Given the many assumptions involved in our model
this agreement is quite remarkable.

Before evaluating the thermal power of our objects we comment on how our
results depend on the mass estimates. For example, an increase of a factor 2
in the black hole mass will result in an increase of a factor $\sim 2^{1/2}$
of the power at the peak of the thermal emission, but the frequency of such
peak will be lowered by roughly the same factor. The total thermal power does
not depend on the mass, because it is basically given by $\epsilon \dot M c^2 = L_{ion}$.

\section{Disentangling the Luminosity Components}

The results reported in the previous section suggest that the method we are
using to evaluate the thermal emission from the accretion disc in FSRQ is
quite reliable.

We then use the previously evaluated masses and accretion rates to compute the
accretion disc spectra. This time we choose $f=1.23$ in eq. (1), a value more
appropriate for rotating black holes (Malkan 1982), a situation appropriate
for radio-loud sources (e.g., Wilson \& Colbert 1995; Moderski, Sikora \&
Lasota 1998). The thermal spectra have been computed for all the 136 sources,
but optical spectra are only available for the 113 objects newly identified by
the DXRBS. For this reason, the remaining 24 FSRQ have not been used in the
following analysis.

The $\lambda$ vs. $f_{\lambda}$ measured optical spectra were converted to
$\nu$ vs. $L_{\nu}$ using our adopted cosmology and were k-corrected using the
optical spectral indices $\alpha_{opt}$ from the DXRBS database.  The computed
thermal spectra were then compared to the observed ones.

An interesting quantity to measure is the ratio of thermal to total power at a
fixed rest-frame frequency. Unfortunately, the spread in redshift of the
sources makes it impossible to evaluate this ratio at the same frequency for
all objects. We then selected three rest-frame frequency intervals around the
line-free continuum zones discussed by Forster et al. (2001) to evaluate the
ratio $L_{th}/L_{tot}$ in these bands for the FSRQ sample. Namely, the
interval $7.08 \times 10^{14} - 7.24 \times 10^{14}$ Hz is covered by 54
objects, the $1.32 \times 10^{15} - 1.35 \times 10^{15}$ Hz one is covered by
76 sources, while 27 spectra cover the interval $2.24 \times 10^{15} - 2.29
\times 10^{15}$ Hz. $L_{th}$ was evaluated at the central frequency of the
interval, while $L_{tot}$ was obtained by taking the mean power in the
interval. In cases where the frequency interval fell into the noise or
included strong absorption lines the ratio was not determined.

\begin{figure}%3
\epsfysize=9cm 
\hspace{3.5cm}\epsfbox{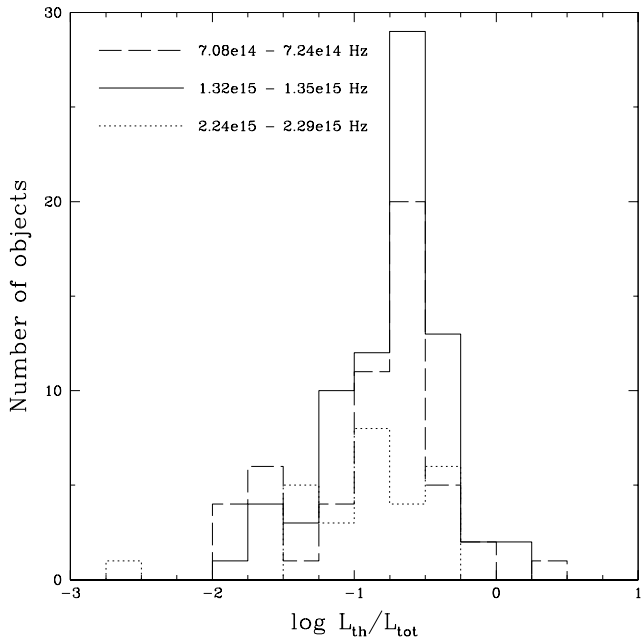}
\caption[h]{The distribution of the ratios of thermal to total luminosity in
three line-free continuum intervals: $7.08 \times 10^{14} - 7.24 \times
10^{14}$ Hz (dashed line), $1.32 \times 10^{15} - 1.35 \times 10^{15}$ Hz
(solid line), $2.24 \times 10^{15} - 2.29 \times 10^{15}$ Hz (dotted line).
\label{hist_lthltot}}
\end{figure}

Table 1 reports the ratio $L_{th}/L_{tot}$ and $L_{tot}$ for the three
frequency bands for the individual sources. Table 3 gives mean and median
values for $\log L_{th}/L_{tot}$ for all three bands, the percentage of
objects with $L_{th}/L_{tot} < 0.5$ (i.e., thermal to non-thermal power ratio
$<1$) per band, and the mean redshift of the sources. Note that for some
sources two of the selected intervals fall within the observed spectrum so two
ratios are available. Fig. \ref{hist_lthltot} shows the distribution of $\log
L_{th}/L_{tot}$ in the three bands.

Table 3 and Fig. \ref{hist_lthltot} show that, typically, the non-thermal
component dominates, as the thermal one makes up $\sim 15$ per cent of the
optical/UV power. Moreover, $\ga 91$ per cent of the $L_{th}/L_{tot}$ values
are below 0.5, which means that in only $\la 9$ per cent of the cases the
thermal power is larger than the non-thermal one. Note that $L_{th}/L_{tot}$
is $> 1$, an unphysical result, only in 3 cases ($\sim 2-3$ per cent). This is
another confirmation of the reliability of our method.

The spectra of the majority ($\sim 70$ per cent) of our FSRQ have not been
observed at parallactic angle, which results in a flux loss due to atmospheric
differential refraction, more pronounced in the blue part of the spectrum. We
have nevertheless included all the sources in our study, based on three facts:
1. the loss at the position of a given line, as tabulated by Filippenko
(1982), is $\ga 30$ per cent only in $\sim 10$ per cent of the cases; this is
due to the fact that most observations were done with the source at meridian;
2. the $L_{th}/L_{tot}$ ratio for sources observed at parallactic angle is not
significantly different from that for sources not observed at parallactic
angle. In particular, in the $1.32 \times 10^{15} - 1.35 \times 10^{15}$ Hz
range, which is the most populated one, $L_{th}/L_{tot}$ for the parallactic
angle sources is $\sim 30$ per cent smaller than for all sources; if anything,
then, this would go in the direction of decreasing the thermal component;
3. the parameter which enters more prominently in our calculations is
$L_{BLR}$, which is in most cases based on more than one line. The effect of
the parallactic angle, therefore, gets diluted. Slit sizes during our
observations were matched to the seeing. We are not aware of any other
systematic effects which could have influenced our flux calibration.

\begin{table*}
\centerline{\bf Table 3. Thermal to Total Power Ratios}
\begin{center}
\begin{tabular}{ccccrc}
\hline
Band & \multicolumn{2}{c}{$\log L_{th}/L_{tot}$}& N & Percentage $<0.5$ & Mean redshift \\
     & Mean & Median &   &                 &               \\
\hline
$(7.08 - 7.24) \times 10^{14}$ Hz & $-0.85\pm0.07$ & $-0.73$ & 54 & 91 & $0.67\pm0.04$ \\ 
$(1.32 - 1.35) \times 10^{15}$ Hz & $-0.75\pm0.05$ & $-0.70$ & 76 & 93 & $1.43\pm0.07$ \\
$(2.24 - 2.29) \times 10^{15}$ Hz & $-0.92\pm0.09$ & $-0.85$ & 27 &100 & $2.32\pm0.10$ \\
\hline
\end{tabular}
\end{center}
\end{table*}

We now use the luminosity values to perform correlations between powers. Two
possible complications should be kept in mind: first, when correlating
luminosities one has to be particularly careful to remove the (generally)
common redshift dependence which can introduce spurious correlations. The
proper way of dealing with the problem is to examine the correlations between
luminosities excluding the dependence on redshift, i.e., via a partial
correlation analysis (see Kendall \& Stuart 1979; Anderson 1984);
second, the three frequency ranges sample different redshifts, as shown in
Tab. 3, and therefore the comparison is subject to possible evolutionary
effects.

\begin{figure} %4
\epsfysize=9cm 
\hspace{3.5cm}\epsfbox{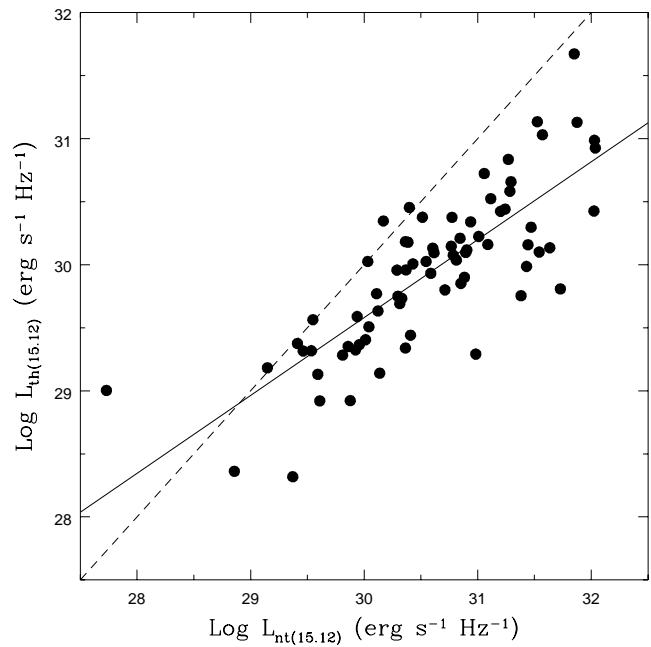}
\caption[h]{The thermal vs. non-thermal luminosity relation at $\log \nu
=15.12Hz$ (rest-frame) for our FSRQ. The solid line is our best fit to the
data: $\log L_{th} = 0.62 \log L_{nt}+11.05$; the dashed line represents the
locus of $L_{th}=L_{nt}$.\label{lth_lnt}}
\end{figure}  

We have correlated the thermal power $L_{th}$ with the non-thermal one,
$L_{nt}$, starting from the $1.32 \times 10^{15} - 1.35 \times 10^{15}$ Hz
range, which includes the largest number of sources. We assume that two
components make up the optical/UV emission in our sources, a thermal one
related to the accretion disc, and a non-thermal one related to the jet. The
former has then been derived by subtracting the thermal from the total power,
i.e., $L_{nt} = L_{tot} - L_{th}$. Fig.  \ref{lth_lnt} shows that the two
quantities are strongly correlated ($P>99.99$ per cent, $R=0.80$).  This is
true even when the common redshift dependence is subtracted off. Note that the
observed strong correlation cannot be due to the fact that the two powers are
not strictly independent, as that would result in an {\it anti-correlation}
between the two quantities. Both powers are also strongly correlated with
optical power $L_o$ (defined at 5,000 \AA) but, again, their mutual
correlation remains highly significant ($P>99.9$ per cent) even when this
common dependence is taken into account. (We note that the $L_{nt} - L_o$
correlation is expected, given that $L_{nt} \approx L_o$.) The fact that
$L_{th} \propto L_{nt}^{0.62\pm0.05}$ suggests that the $L_{th}/L_{tot}$ ratio
depends on luminosity. Indeed, we find a strong anti-correlation between
$L_{th}/L_{tot}$ and $L_{nt}$ (but not $L_{th}$), i.e., the stronger the
non-thermal power, the lower the thermal fraction, with this ratio approaching
unity at the low end of our luminosity range (see Fig. \ref{lth_lnt}). This
could indicate that only in less powerful (possibly less beamed?) sources the
thermal component manages to rival with the non-thermal one. The fact that
weak-lined blazars, which are normally less powerful than strong-lined ones,
have a very weak, if any, thermal component, and therefore lie significantly
off the correlation, would then suggest a dichotomy between the two classes.

As regards the other two bands, we also find a slope less than 1 for the
correlation between the two powers, namely $L_{th} \propto
L_{nt}^{0.74\pm0.11}$ and $L_{th} \propto L_{nt}^{0.53\pm0.15}$ for
$7.08\times 10^{14} - 7.24\times 10^{14}$ Hz range and the $2.24\times 10^{15}
- 2.29\times 10^{15}$ Hz range respectively.  In both cases the significance
of the $L_{th} - L_{nt}$ correlation is very high ($99.8$ and 99.7 per cent)
once the redshift dependence is subtracted off but becomes low ($P \sim 78$
and 92 per cent) once the effect of the optical power is also considered.

\section{The Disc and Jet Powers}\label{disc_jet}

In the last section we evaluated the thermal component in the optical/UV band,
subtracted it from the total power and computed the ratio $L_{th}/L_{nt}$ of
monochromatic luminosities for some fixed frequencies. This led us to
recognize that our sources' optical/UV spectra are almost all dominated by
non-thermal emission. However, we still do not know which of the two outputs
dominates the {\it total} power. This is because: 1) the non-thermal emission
is amplified by the beaming effect; 2) the total radiative power is only a
fraction of the total power in the jet; 3) the peak of the thermal luminosity
(located in the UV - soft X-ray band) does not fall in the region covered by
the available spectra, and thus thermal emission may overcome the non-thermal
one (which is usually described by $L_{nt}(\nu) \propto \nu^{-\alpha}$,
$\alpha > 0$) at higher frequencies. 

For this reason we want to compare directly the total powers from the disc and
jet; in the following these powers are referred to as $L_{Th}$ and $L_{Jet}$, with 
the first letter of each subscript in upper case, to indicate that these quantities are
powers integrated over the whole spectral range and not luminosities per unit of frequency 
as in the previous section. 
$L_{Th}$ can be easily calculated by simple integration over
the thermal spectra previously computed and is also basically equal to the
ionizing power $L_{ion}$. The latter one is not easy to evaluate, however,
because of beaming effects. We follow Willott et al. (1999) and define the
time-averaged jet power as the total energy transported from the central
engine by both jets divided by the age of the radio source. Willott et
al. (1999) have derived this relationship to evaluate the jet power from the
monochromatic luminosity at 151 MHz,

$$L_{Jet} \simeq 3\,\,10^{38}\,\,c\,\,L_{151}^{6/7}\,\,\,W, \eqno (10)$$ 
where $L_{151}$ is in units of $10^{28}$ W Hz$^{-1}$ sr$^{-1}$ and $c$ is a
factor which takes into account all the uncertainties of the calculation and
lies in the range $\sim 1 - 20$. In particular, these uncertainties reside in:
(i) the volume filling factor of the synchrotron emitting material; (ii) the
contribution to the energy budget from protons; (iii) the dimension of the
reservoir of mildly relativistic electrons; (iv) the departure from the
minimum-energy condition in the lobe material. Relation (10) holds if
$L_{151}$ is the unbeamed radio luminosity, that is, the extended component of
the emission at 151 MHz. As we only have measurements for the radio luminosity
at 5 GHz, to estimate the jet power we need to extrapolate to 151 MHz.

The unbeamed power at 151 MHz can be expressed as follows:

$$L_{151,ub} = L_{5,ub}\,\left({151\over 5000} \right)^{-\alpha_{ub}} 
\equiv {L_{5,b}\over R_{5}}\,\left({151\over 5000} \right)^{-\alpha_{ub}};
\eqno (11)$$ 
where $\alpha_{ub}$ is the spectral index of the unbeamed component, $L_{5,b}$
and $L_{5,ub}$ are the 5 GHz powers of the beamed and unbeamed component,
respectively, and $R_{5}$ is the core dominance parameter at 5 GHz.

We do not have $R_{5}$ values for all our sources. However, as the radio
spectral index is a beaming indicator, we expect a relation to hold between
$R_{5}$ and $\alpha_r$. To derive this relation we use all FSRQ and SSRQ
(steep spectrum radio quasars) in the DXRBS database for which both $R_{5}$
and $\alpha_r$ are available (65 sources). The correlation we find is the
following:

$$\log R_{5}=-(0.65\pm0.16)\,\alpha_b +0.84,\eqno (12)$$
significant at the $>99.99$ per cent level. As discussed in Landt et al.
(2001), the DXRBS core-dominance parameters, based on Australia
Telescope Compact Array (ATCA) data, are actually upper limits. Indeed, the
same correlation for a (highly inhomogeneous) set of 49 radio-loud quasars
with higher resolution radio data in the multiwavelength AGN catalog of
Padovani, Giommi \& Fiore (1997) yields a a best-fit correlation with a slope
of $-0.88\pm0.21$, consistent within the errors with ours, but with a lower
intercept value. The best-fit value at $\alpha_r = 0.18$, the mean for the
DXRBS sample, is in fact a factor $\sim 4$ smaller. In the following we then
adopt the relation given by eq. (12) but with an intercept of 0.24, so that
our slope comes from an homogeneous set of data. Note that this revised
correlation gives a dividing value between core-dominated ($R_{5} > 1$) and
lobe-dominated ($R_{5} < 1$) sources of $\alpha_r \sim 0.4$, not too far from
the generally adopted distinction between FSRQ and SSRQ ($\alpha_r = 0.5$).

Thus, assuming $\alpha_{ub}=0.8$, a typical value for the radio spectral index
of the Fanaroff-Riley II (FR II) radio galaxies, and since $L_{5,b}/R_{5} =
L_{5}/(R_{5}+1)$, we can evaluate $L_{151,ub}$. Namely:

$$L_{151,ub} = {L_{5} \left({151\over 5000} \right)^{-0.8}\over 
(10^{-0.65\,\alpha_r +0.24}+1)}.\eqno (13)$$

\begin{figure} %5
\epsfysize=9cm 
\hspace{3.5cm}\epsfbox{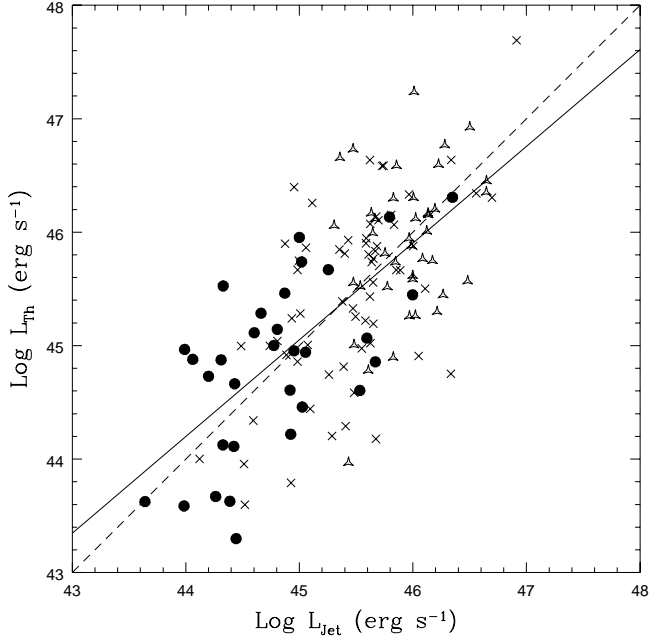}
\caption[h]{The relation between thermal and jet power using $c=20$ in
eq. (10). Filled circles (open triangles) represent sources whose mass has
been evaluated using the H$\beta$ (C IV) line, and crosses are objects whose
mass has been inferred from the correlation shown in Fig. \ref{mass_lion}. The
solid line represents our best fit to the data; the dashed line represents the
locus of $L_{Jet}=L_{Th}$.\label{lth_ljet_20}}
\end{figure}

Figure \ref{lth_ljet_20} plots $L_{Th}$ vs. $L_{Jet}$, with $L_{Jet}$
calculated using $c=20$ in eq. (10). 
We find that the two powers are correlated, albeit with a large
scatter. Namely, $\log L_{Th} = 0.85\,\log L_{Jet} + 6.69$ (the intercept
becomes 7.80 if $c=1$), $R=0.67$, $P>99.99$ per cent. The correlation still
holds if the common redshift dependence is subtracted off ($R=0.40$, $P=99.99$
per cent).

\begin{figure}%6
\epsfysize=9cm 
\hspace{3.5cm}\epsfbox{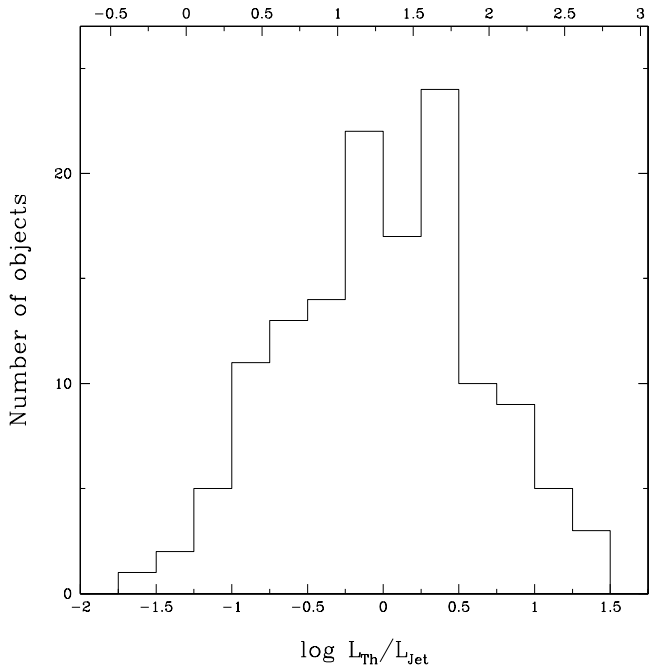} 
\caption[h]{The ratio of thermal to jet power for our sample. The bottom axis 
gives the range using $c=20$ in eq. (10), while the top axis shows the range
for $c=1$.\label{disk_jet_ratio}}
\end{figure}

Figure \ref{disk_jet_ratio} shows the distribution of the thermal to jet
power, $\log L_{Th}/L_{Jet}$, for our sample using $c=20$ in eq. (10) (bottom
axis) and $c=1$ (top axis). The mean value for the two cases are,
respectively, $-0.01\pm0.05$ and $1.29\pm0.05$ or $\langle L_{Th}/L_{Jet}
\rangle \sim 1$ and 20.

As a check of our method for evaluating the bolometric thermal output from the
accretion disc we have used another relation in Willott et al. (1999), namely
$L_{Th} \simeq 5\times 10^3 L_{\rm \OII}$, where $L_{\rm \OII}$ is the power
of the \OII~narrow line. This is based on assuming an equivalent width of 10
\AA~for \OII, which allows an estimate of the blue luminosity, which is then
converted to bolometric, under the assumption that the bolometric output is
dominated by the the accretion disc emission. We have estimated $L_{\rm \OII}$
for our sources using the correlation with $L_{151}$ found by Willott et
al. (1999) for FR II radio sources. We find that the thermal bolometric
luminosity evaluated from the relations of Willott et al. (1999), $L_{Th,2}$,
is very well correlated ($R \sim 0.6$, $P \sim 99.99$ per cent) with that
derived with our method, $L_{Th,1}$, albeit with a very large scatter, with a
mean ordinary least-squares (OLS) slope (Isobe et al. 1990) of 1.0. The mean
values for the two powers are $\langle \log L_{Th,1} \rangle = 45.44 \pm 0.07$
and $\langle \log L_{Th,2} \rangle = 45.34 \pm 0.06$. The two means and
distributions are not significantly different according to a Student's t-test
and a KS test respectively. We note that our method is more robust as we
employ more than one emission line and no assumption on the equivalent width
is made. This comparison, nevertheless, provides independent evidence that the
method we use to compute the bolometric thermal power is reliable.

\begin{figure} %7
\epsfysize=9cm 
\hspace{3.5cm}\epsfbox{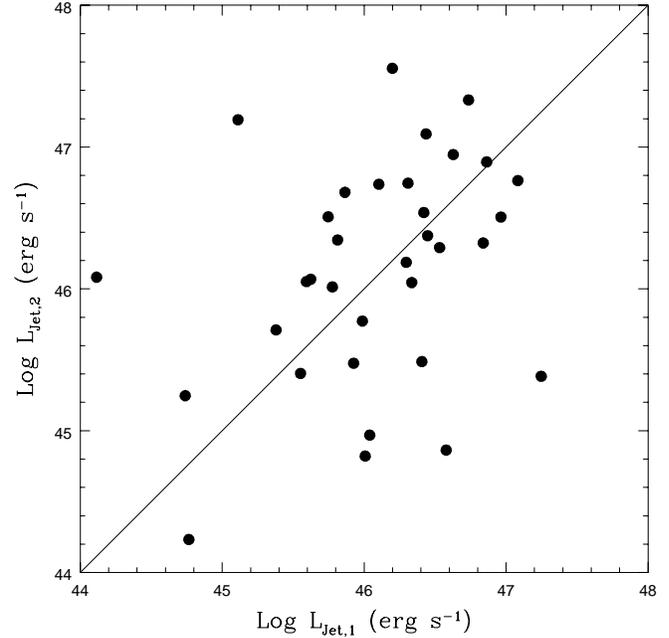} 
\caption[h]{The comparison of the jet powers published in Celotti et
al. (1997) ($L_{Jet,2}$) with those computed with eq. (10) assuming $c=20$
($L_{Jet,1}$) for the 34 FSRQ studied by these authors. The solid line
represents the locus of $L_{Jet,1}=L_{Jet,2}$.\label{ljet2_ljet1}}
\end{figure}

Finally, we want an independent method to evaluate the jet power, for
comparison with eq. (10). Celotti et al. (1997) give the jet power for 34
objects (19 high polarized, core dominated quasars and 15 low polarized, core
dominated quasars), based on a synchrotron self-Compton model (SSC), applied
to the radio (Very Large Baseline Interferometry [VLBI]) and X-ray fluxes.
Their core dominance parameter $R_{5}$ can be found in Ghisellini et
al. (1993). We then collected from the literature the luminosities at $5$ GHz
and used eq. (11) with $\alpha_{ub}=0.8$ to compute the jet power
and make a comparison. The results are shown in Fig. \ref{ljet2_ljet1}, where
$L_{Jet,1}$ is the jet power evaluated using eq. (10) with $c=20$ and
$L_{Jet,2}$ is the jet power given in Celotti et al. (1997). We note that,
despite the large scatter, the two powers are correlated ($P\sim 95$ per
cent), with a mean OLS slope of $0.94$. The mean values of the two sets of
measurements are $\langle \log L_{Jet,1} \rangle = 46.1 \pm 0.1$ and $\langle
\log L_{Jet,2} \rangle = 46.1 \pm 0.1$, i.e., the maximum powers evaluated
with eq. (10) are consistent with those published in Celotti et
al. (1997). This good agreement favours a large value of $c$.

The study of the underlying correlations between $L_{Th}$, $L_{Jet}$, and
their ratio with effective spectral indices (radio-optical $\alpha_{\rm ro}$,
optical-X-ray $\alpha_{\rm ox}$, and radio-X-ray $\alpha_{\rm rx}$, defined
between the rest-frame frequencies of 5 GHz, 5,000 \AA, and 1 keV) and powers
is complicated by the very strong $L_{Jet} - L_{r}$ correlation which is built
in due to the way we derived $L_{Jet}$. We studied the correlation matrix of
$L_{Th}, L_{Jet}, L_r, L_o, L_x$ plus redshift to assess which ones are the
strongest correlations. Note that, whenever X-ray flux is involved, we have
excluded sources with {\it ROSAT} Position Sensitive Proportional Counter
(PSPC) offsets between $13^{\prime}$ and $24^{\prime}$. This area of the PSPC,
in fact, is affected by the supporting structure of the instrument, which
attenuates the X-ray flux. We do not address here correlations between radio,
optical, and X-ray powers. The only significant primary (non-trivial)
correlations are $L_{Th} - L_o$ and $L_{Th} - L_x$. All correlations involving
$L_{Jet}$ are secondary and due to the $L_{Jet} - L_r$ correlation. No
correlations are present between $L_{Jet}$ and the effective spectral indices
once the effect of $L_r$ is taken into account. On the other hand, $L_{Th}$
correlates with $\alpha_{\rm ro}$ (negatively) and $\alpha_{\rm ox}$
(positively). The $L_{Th}/L_{Jet}$ ratio anti-correlates strongly with
$\alpha_{\rm ro}$ and $\alpha_{\rm rx}$, but the latter correlation appears to
be induced by the former.

To conclude, we now discuss how our results would change for different choices
of covering factor and black hole. If a covering factor of $\sim 30$ per cent
instead of $10$ per cent is assumed all the computed thermal powers decrease
roughly by a factor $\sim 3$. The dominance of the non-thermal emission in the
optical/UV would become even stronger, as the thermal component would only
make up $\sim 7$ per cent of the power. The average ratio between the total
disc and the jet power would be $\sim 0.3$ ($\sim 7$) for $c=20$ ($c=1$). In
this case, our thermal powers would be on average a factor $\sim 2$ below
those estimated using the Willott et al. correlations. Assuming a non-rotating
black hole would increase the accretion rates by a factor $\sim 5$, as the
efficiency goes down but the ionizing power is fixed by the BLR power (see
eq. 8). The shape of the spectrum also changes because the radius of the last
stable orbit increases. The combination of these two effects result in a
thermal spectrum which is not too different from that of a rotating black
hole, with a slight shift towards lower frequencies and a slightly higher
normalization. 

\section{Discussion}\label{discussion} 

Three main results have come out of this work. We discuss them now in turn. 

\subsection{The Thermal Component in the Optical/UV Spectra of Flat-spectrum Radio 
Quasars}

It has long been thought that any thermal emission related to accretion in the
optical/UV spectra of FSRQ had to be relatively weak, or rather swamped by the
non-thermal emission from the jet (although exceptions exist, e.g., 3C 273).
We have quantified this in a systematic, albeit model-dependent way, by
deriving an accretion disc spectrum and by then comparing it to the observed
rest-frame optical/UV spectra of more than 100 FSRQ from the DXRBS. The
determination of the mass and accretion rate for our sources, which are
required parameters of the model, should be relatively robust. The former is
based on the virial theorem, by now widely used, normalized on the results of
reverberation mapping campaigns. (Even the mass determinations based on C IV
agree very well with a recent calibration published by Vestergaard (2002).)
The latter has been derived from the BLR power and the only uncertainty there
is the covering factor, which we think should be constrained within a factor
$\sim 3$. Moreover, the reliability of our method has been tested by comparing
the predictions of the model with the observed optical/UV spectra of a sample
of radio-quiet quasars, under the assumption that in this case most of the
emission should be due to the disc. Reassuringly, the predictions were within
a factor of two from the observations. 

We have found that the optical/UV emission of FSRQ is dominated by a
non-thermal component, which makes up $\sim 85$ per cent of the total
power. Only in a small fraction ($\la 9$ per cent) of the objects does the
thermal component dominates. This is consistent with the relatively steep
optical/UV slopes of DXRBS FSRQ ($\alpha \sim 1.2$; Landt et al., in
preparation). In all three rest-frame bands we have used there is evidence
that the thermal fraction is higher at lower non-thermal powers, which might
suggest that only in less powerful, and therefore possibly less beamed,
sources the thermal component manages to rival with the non-thermal one. 

Whiting, Webster \& Francis (2001) have fitted the optical and near-infrared
SEDs of $\sim 100$ FSRQ from the Parkes Half-Jansky flat-spectrum sample.
They assessed the relevance of thermal and non-thermal components by fitting a
blue power law (fixed to $\alpha_B = 0.3$) and optical synchrotron emission.
The ratio of the synchrotron and power-law components at a rest-frame
wavelength of $5,000$ \AA~was found to span nearly four orders of
magnitude. The distribution was strongly peaked, however, at ratios $\ga 10$,
suggesting a dominance of non-thermal emission in the optical band of these
sources, in agreement with our results obtained with an independent method.

\subsection{The Masses of Radio-loud Quasars}

It has been suggested in the literature that radio-loud quasars require
massive black holes. Laor (2000) pointed out that nearly all Palomar-Green
(PG) quasars with $M > 10^9 M_{\odot}$ are radio-loud, while quasars with $M <
3 \times 10^8 M_{\odot}$ are practically all radio-quiet. This has been
recently taken up by Boroson (2002), who has suggested a scheme to explain the
radio-loud/radio-quiet dichotomy based on mass and Eddington ratio. Ho (2002),
however, by using a sample which spanned a large range of activity, did not
confirm the trend found by Laor (2000). Lacy et al. (2001) also found a
dependence of black hole mass on the radio-loudness parameter $R^*$
(equivalent to our $\alpha_{\rm ro}$) with most radio-loud ($R^* > 10$)
sources having $M > {\rm a~few} \times 10^8 M_{\odot}$, but with the
radio-quiet ones extending to $\ga 10^9 M_{\odot}$.

About 25 per cent of our sources, which are all strongly radio-loud ($\langle
\alpha_{\rm ro} \rangle \sim 0.65$, well above the typically assumed dividing
line of 0.2), have $M < 3 \times 10^8 M_{\odot}$. If we restrict ourselves to
the objects whose mass has been derived from $H\beta$, for comparison with
previous studies, this fraction gets much higher. The mean mass in this case
is $\langle \log M \rangle = 8.41\pm0.09\, M_{\odot}$ (to be compared with
$\langle \log M \rangle = 8.74\pm0.04\, M_{\odot}$ for the whole sample) and 
now $\sim 47$ per cent of our objects have estimated black hole masses below
$3 \times 10^8 M_{\odot}$. Keeping in mind the danger of comparing results for
samples selected with different criteria, the mass range covered by our
radio-loud quasars is not very different from that of the radio-quiet quasars
in Laor (2000).

If the motion of emission-line gas is confined predominantly to a disc
perpendicular to the radio axis then in core-dominated (flat-spectrum)
sources, like ours, we might be biased towards low FWHM values (Wills \&
Browne 1986) and, therefore, we might be underestimating the black hole
masses. Although this could have some effect on our results, we believe it
cannot be major (see also Oshlack et al. 2002). First, within DXRBS we do not
find any correlation between FWHM($H\beta$) (and also FWHM(C IV)) and
$\alpha_{\rm r}$, although the relative number of SSRQ in DXRBS is, by
construction, small ($\sim 24$ per cent and $\sim 18$ per cent for sources
with $H\beta$ and C IV respectively), as we have a cut at $\alpha_{\rm r} \le
0.7$. We also do not find any correlation between ATCA core-dominance
parameter and FWHM($H\beta$) (20 sources) or FWHM(C IV) (27 sources). Second,
Gu, Cao \& Jiang (2001) have assembled FWHM($H\beta$) for 86 radio sources,
selected from the 1 Jy, S4, and S5 catalogs, including 55 FSRQ and 31
SSRQ. Table 2 in Oshlack et al. (2002) shows that the ratio of the mean
(median) FWHM($H\beta$) for SSRQ and FSRQ is $\sim 1.44$ ($\sim 1.25$), which
would imply a ``correction'' factor for our $H\beta$ masses $\sim 2$ ($\sim
1.6$). This means that $\sim 28$ per cent of our low-redshift objects would
still have estimated black hole masses below $3 \times 10^8 M_{\odot}$. As
discussed in Sect. 4, a factor of 2 increase in black hole mass has in any
case a small effect on our other results.

Oshlack et al. (2002) have estimated black hole masses for 39 sources
belonging to the Parkes Half-Jansky flat-spectrum sample, also finding that a
significant fraction of the radio quasars in their sample have relatively
small masses. Both studies are based on complete, radio-selected samples so
the failure of some previous studies to find radio-loud quasars with small
masses is likely due to a selection effect. Oshlack et al. (2002) comment on
the fact that there is a strong correlation between mass and optical/UV
luminosity, shown for our sample in Fig. \ref{mass_lion}, a consequence of
using ionizing power to estimate the BLR radius. This implies that if a sample
is biased against selecting low optical luminosity, radio-loud quasars, as
seems to be the case for the PG sample, it will miss the low-mass radio-loud
sources. DXRBS is almost completely identified and we find a large range of
optical powers in our quasar sample. 

Contrary to what found by Laor (2000), we find an anti-correlation between
mass and $\alpha_{\rm ro}$, in the sense that the most radio-loud sources have
also the smaller masses. Again, this is partly induced by the correlations
between mass and ionizing (optical) power and $\alpha_{\rm ro}$ and optical
power (see also Oshlack et al. 2002).

The relationship between black hole mass and radio power has also been recently
addressed in the literature. Since all our sources have a radio flat-spectrum
and are therefore heavily affected by relativistic beaming, we cannot comment on
this point.

\subsection{The Disc - Jet Connection in Flat-spectrum Radio Quasars}

Even more interesting than the non-thermal/thermal ratio in the optical/UV
band is the relationship between integrated disc and jet emission. The former
is easily derived assuming that the ionizing photons originate in the
accretion disc and therefore depends on the BLR luminosity and the covering
factor. Note that the total disc emission is independent of the mass. An
estimate of the total jet power is more difficult to obtain, because it
depends on many astrophysical parameters, some of which are at present not
well constrained. We have then relied on the theoretical relationship obtained
by Willott et al. (1999), which links jet power to the extended radio emission
at 151 MHz, derived for our sources from their 5 GHz radio power and radio
spectral index. This relationship has a built-in uncertainty $\sim 20$, due to
various uncertainties in the model assumptions.

We compared the results of both estimates to previous results obtained using
independent methods, with relatively good agreement. In particular, an
application of our method to estimate jet power to the 34 quasars studied by
Celotti et al. (1997) using an SSC model favours large values of $c$ and would
actually suggest a nominal value $c \sim 20$.

We have found that $L_{Th} \propto L_{Jet}^{0.9}$ and that, on average,
$L_{Th}/L_{Jet} \sim 1$ (or 20 if $c=1$). The dispersion around the average
is relatively large, about a factor of 4 ($1\sigma$). This could simply
reflect the uncertainties of our method or might indicate a dependence of this
ratio on some parameter. One of the strongest correlations we find is between
$L_{Th}/L_{Jet}$ and the optical-to-radio flux ratio, which we parametrize
using $\alpha_{\rm ro}$. This correlation might be expected given that $L_{Th}
\propto L_{o}$ and $L_{Jet} \propto L_{r}$. (We note however that we have
shown that $L_o$ is dominated by the non-thermal component, while $L_{Jet}$
depends by construction on the extended radio emission, while our sources are
mostly core-dominated.) However, it is not induced by other correlations and
therefore might be used to estimate the total disc to jet power ratio as a
function of radio-loudness.

\begin{figure}   %8
\epsfysize=9cm 
\hspace{3.5cm}\epsfbox{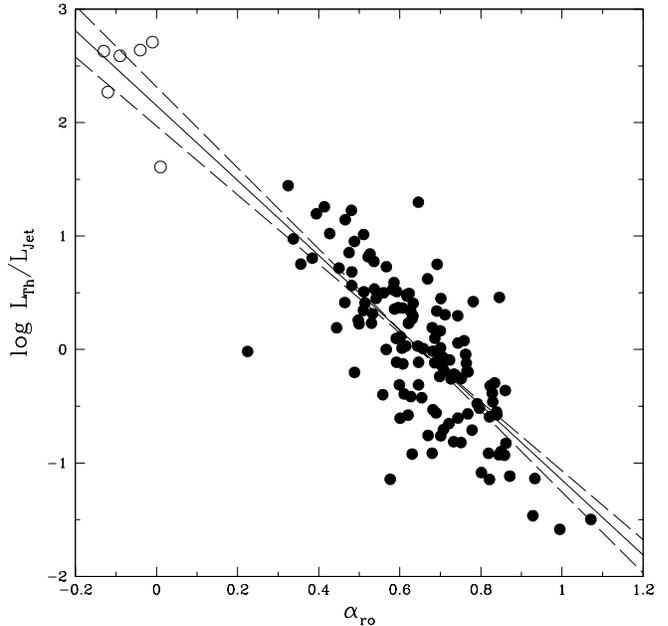} 
\caption[h]{The ratio of thermal to jet power (using $c=20$ in eq. 10)
vs. $\alpha_{\rm ro}$. Filled circles represent our FSRQ while open
circles represent the subset of radio-quite quasars in Tab. 2 with
extended radio power information. The solid line represents the best
fit to the FSRQ data, while the dashed lines indicate the $1\sigma$
dispersion. See text for details. 
\label{disk_jet_aro}}
\end{figure}

Fig. \ref{disk_jet_aro} plots $L_{Th}/L_{Jet}$ (using $c=20$ in eq. 10)
vs. $\alpha_{\rm ro}$ for our sources. The best fit is $\log L_{Th}/L_{Jet} =
-(3.30\pm 0.26) \alpha_{\rm ro} + 2.15$. (The intercept value becomes 3.45 if
$c=1$ is used instead.) Assuming that this relationship can be extrapolated to
the radio-quiet regime, we get $L_{Th}/L_{Jet} \sim 200$ for $\alpha_{\rm ro}
\sim -0.05$, a value typical for radio-quiet AGN. (If $c=1$ we get instead
$L_{Th}/L_{Jet} \sim 4,000$.) We checked this extrapolation by estimating the
jet power for the radio-quiet (PG) quasars used in Sect. 4 to test our
accretion disc model. Namely, we derived the extended radio flux at 5 GHz for
the sources in Tab. 2 by subtracting the core from the total fluxes given in
Kellermann et al. (1989). Only the six sources for which this difference was
larger than the quoted noise summed in quadrature (0.09 mJy) were
considered. The flux was then extrapolated at 151 MHz assuming a radio
spectral index of 0.8 and then eq. (10) was used to estimate the jet
power. The radio-quiet quasars fall on the extrapolation of the relation
obtained for the FSRQ in the $L_{Th}/L_{Jet} - \alpha_{\rm ro}$ plane. This
supports the use of $\alpha_{\rm ro}$ to estimate the disc to jet power ratios
in all broad-line AGN and suggests that the disc to jet power ratio in
radio-quiet AGN is a few hundred times larger than in radio-loud AGN. The disc
power for the two classes, at least for the sources discussed in this paper,
is comparable, so what is changing is obviously jet power.

Note that in both classes of sources the contribution of the host galaxy to
the optical emission (and therefore to $\alpha_{\rm ro}$) is negligible. The
DXRBS FSRQ are typically at $z \sim 1.5$, where the strong nuclear luminosity
evolution should make the AGN dwarf the host galaxy. As regards the PG
quasars, these were selected on the basis of being ``point-like''. Detailed
imaging studies (e.g., McLeod \& McLeod 2001) have indeed confirmed that the
ratio of host to nuclear light is $\ll 1$ in the near-infrared band, and
therefore even smaller in the optical. 

\subsubsection{Comparison with Previous Results}

We now want to compare our results with previous works also aimed at studying
the relation between thermal and jet power in radio-loud sources.

Rawlings \& Saunders (1991) found a very strong correlation, covering four
orders of magnitude, between the jet kinetic power and narrow line region
luminosity in radio-loud objects. Since the power of the narrow line region is
proportional to the thermal power, this translates in a correlation between
the latter and the jet power, similar to ours. The slope of our correlation is
close to that of Rawlings \& Saunders (1991), and their average ratio between
thermal and jet power is close to unity (we find an average value of $\sim
1$, for $c=20$), even if their computation of the kinetic power is carried
out in a different way, i.e., by invoking equipartition between the magnetic
field and the electron energy densities in the jets.

Celotti et al. (1997) found a weak (95.8 per cent) correlation 
between the jet kinetic energy and the power of the broad line
region for an inhomogeneous sample of radio-loud sources composed of Steep
Spectrum Radio Quasars, FSRQ, and BL Lacertae objects, with the thermal and
jet powers of the same order of magnitude. 

Willott et al. (1999) evaluated the relation between the thermal and
non-thermal power for a very large sample of FR II radio sources. We have
adopted their method to estimate jet power but they evaluated the thermal
power from the \OII~line luminosity. They found an average thermal to jet
power ratio $\sim 1 - 2$ for $c=20$ depending on the cosmology adopted ($q_0 =
0.5$ and $q_0=0$, respectively) and $20 - 40$ for $c=1$. The values we find
for our FSRQ sample are $\sim 1$ and $20$ (see also
Fig. \ref{disk_jet_ratio}), consistent within a factor of two. Thus, we can
conclude that our blazars behave as their parent population, the FR II radio
galaxies, and this is definitely in favour of unified scheme models (Scheuer
1987; Barthel 1989; Antonucci 1993; Urry \& Padovani 1995).

Maraschi \& Tavecchio (2002) estimated the jet power for 11 FSRQ using
synchrotron inverse-Compton fits to multiwavelength data and the accretion
power for the same objects using the optical-UV bump in their data or the BLR
luminosity when the data did not feature such a bump. In agreement with
Celotti et al. (1997), they found that the two powers are of the same order of
magnitude.

\subsubsection{Jet Models}

The leading model to explain jet production is the Blandford \& Znajek (1977;
BZ) mechanism, that involves the electrodynamic extraction of rotational
energy from a spinning black hole (i.e., a Kerr hole). The BZ power scales as
$$L_{BZ} \propto r^2B^2,\eqno(14)$$
$r$ being the radius of the Kerr hole and $B$ being the magnetic field
threaded to the hole (see also Cavaliere \& D'Elia 2002 for a detailed
discussion). However, the BZ mechanism has been proven to be insufficient to
achieve the luminosities of the most powerful FSRQ due to the magnetic field,
which is limited by the radiation pressure in the accretion disk. Ghosh \&
Abramowicz (1997), for example, have shown that in a radiation pressure
dominated disk, the {\it maximum} value of the electromagnetic energy that can
be extracted from the Kerr hole is
$$L_{K} \approx 2\;10^{45}\;M_9\;a^2\;\;\;erg\;s^{-1},\eqno(15)$$
where $M_9$ is the Kerr hole mass in units of $10^9 M_{\odot}$ and the
parameter $a$ describes the rotation of the hole ($a=1$ for a maximally
spinning hole and $a=0$ for a non-rotating black hole). We can see from
eq. (15) that the jet power in our most luminous FSRQ, which can reach almost
$10^{47}$ erg s$^{-1}$ (see Fig. \ref{lth_ljet_20}), cannot be explained using
the BZ mechanism to extract energy from the Kerr hole, even assuming the
highest mass we have estimated for our sources $\sim 8 \times 10^9 M_{\odot}$
(see also Maraschi \& Tavecchio 2002). For our sample we can calculate $L_{K}$
using eq. (15) and the black hole masses we evaluated and compare it with the
jet power computed from eq. (10) with $c=20$. We find that, while $\langle
\log L_{Jet} \rangle = 45.45 \pm 0.06$, $\langle \log L_{K} \rangle = 45.04
\pm 0.04$, i.e., a factor $\sim 3$ smaller. Thus, on average, even the {\it
maximum} Kerr hole luminosity produced via the BZ mechanism is not able to
explain the estimated $L_{Jet}$, although in $\sim 24$ per cent of our objects
$L_K$ is greater than $L_{Jet}$. Note that for a value of $c=1$ we get
$\langle \log L_{Jet} \rangle = 44.14 \pm 0.06$, 8 times {\it smaller} than
$L_K$, with $L_K$ greater than $L_{Jet}$ for $\sim 94$ per cent of the
objects. But in this case our jet power estimates would be in gross
disagreement with previous derivations. The point remains, however, that the
difference we find between jet and maximum BZ power is smaller than that
found, for example, by Maraschi \& Tavecchio (2002; see their Fig. 3). We
attribute this to the fact that DXRBS is sampling FSRQ of lower power than the
``classical'', high-power ones normally studied.
 
To explain the jet power in FSRQ two variants of the BZ process has been
proposed. The first one invokes the electrodynamic extraction of rotational
energy from the accretion disk too (Blandford \& Payne 1982). This basically
corresponds to using a larger value for $r$ in eq. (14). The second variant
argues for strong magnetic fields in the plunging orbit region, where
relativistic effects could enhance such fields to higher values than those
fixed by the radiation pressure in the accretion disk (e.g., Meier
1999). Further investigations are definitely needed in order to shed more
light on the BZ process and on its role in the production of blazar jets.

\section{Summary and Conclusions}

We have presented an indirect approach to investigate the relationship between
accretion disc and jet in strong-lined blazars. As the luminosity produced
from the accretion disc in the optical/UV band cannot be observed directly
because of the beamed jet emission, we have estimated the disc component using
an accretion disc model for 136 flat-spectrum radio quasars. These sources
belong to the Deep X-ray Radio Blazar Survey (DXRBS) and therefore represent a
well-defined, almost completely identified sample. The model requires an
estimate of the mass and accretion rate for each source. These have been
derived from broad line widths, normalizing our BLR distances to the results
of reverberation mapping campaigns, and ionizing powers, the latter derived
from the BLR power, which we derive from the emission line fluxes, mostly from
our own spectra. Our estimate of the thermal component has been tested for a
sample of radio-quiet quasars, for which the thermal component should
dominate. Indeed, the observed powers at the predicted peak of the thermal
emission were within a factor less than two from those computed with our
method.

Our main results can be summarized as follows: 

\begin{enumerate}

\item The application of our method to the observed spectra of our sources has
shown that non-thermal emission dominates the (rest-frame) optical/UV band of
strong-lined blazars, accounting on average for $\sim 85$ per cent of the
power. Only a small fraction ($\la 10$ per cent) of our flat-spectrum radio
quasars shows a dominance of thermal emission in their optical/UV spectra. 

\item The ratio between {\it integrated} disc and jet power is, on average, 
$\ga
1$, possibly reaching typical values as high as 20. This large range is due
to uncertainties in the assumptions that have gone into the relation we have
used to estimate jet power from extended radio power, derived by Willott et
al. (1999). Nevertheless, a comparison with results from an SSC-based method
favours low values for this ratio. We have used the correlation between 
disc-to-jet power ratio and optical-to-radio flux ratio to estimate that the 
former
parameter is a few hundred times larger in radio-quiet AGN than it is in
radio-loud AGN. This is confirmed by estimating the jet power for a handful
of radio-quiet quasars. 

\item A by-product of our study is an estimate of black hole masses for radio
quasars belonging to a well-defined sample. Contrary to some previous studies,
we have found that a sizeable fraction of our sample ($\sim 25$ per cent; 50
per cent of the $H\beta$ sample) has a relatively small ($M < 3 \times 10^8
M_{\odot}$) mass, which argues that radio-loud sources do not require
particularly massive black holes. Eddington ratios are typically $\sim 0.1$.

\item We have calculated the maximum power that can be extracted from a
rotating black-hole in our sources according to the Blandford-Znajek
mechanism, making use of our mass estimates. If the jet power is as high as
inferred by a comparison with independent methods, it can be explained by the
Blandford-Znajek process only in a minority ($\sim 25$ per cent) of our
sources.
\end{enumerate} 

\section*{Acknowledgements}
This work has been partially supported by the STScI (DDRF) grant
D0001.82274. We thank the referee for helpful comments.

\end{document}